# Universal distribution of component frequencies in biological and technological systems

Tin Yau Pang[a,b] and Sergei Maslov[a,*]

a Department of Biosciences, Brookhaven National Laboratory, Upton, NY 11973
b Department of Physics and Astronomy, Stony Brook University, Stony Brook, NY 11794
* Correspondence to: maslov@bnl.gov

## Abstract

Bacterial genomes and large-scale computer software projects both consist of a large number of components (genes or software packages) connected via a network of mutual dependencies. Components can be easily added or removed from individual systems and their usage frequencies vary over many orders of magnitude. We study this frequency distribution in genomes of ~500 bacterial species and in over 2 million of Linux computers and find that in both cases it is described by the same scale-free power law distribution with an additional peak near the tail of the distribution corresponding to nearly universal components. We argue that this is a general property of any modular system with a multi-layered dependency network. We demonstrate that the frequency of a component is positively correlated with its dependency degree given by the total number of upstream components whose operation directly or indirectly depends on the selected component. The observed frequency/dependency degree distributions are reproduced in a simple mathematically tractable model introduced and analyzed in this study.

## Introduction

Individual components of complex interconnected systems are used with vastly different frequencies. Examples include the frequency with which individual genes and their orthologs are encoded in genomes of different species (*1*); the frequency of local installations of individual software packages in multi-component software projects (*2*); broad power law distributions of the frequency of citations, visitations, or other measures of popularity of individual publications, webpages, YouTube videos, Facebook and Twitter pages, etc. (*3–5*)**;** and power law distribution of word usage frequencies in text (*6*).

The explanations of the observed broad distribution of usage frequency (or popularity) of individual components generally fall into two broad categories. The first category invokes random multiplicative processes (*7*, *8*) recently exemplified by the preferential attachment model of growing networks (*9*, *10*). These models recently invoked to explain frequency distribution of genes in pan-genomes of bacterial species (*11*) by and large ignore functional differences between components so that the ultimate popularity of a component is determined mostly by its age as well as random events in early phases of growth of the system. The second category of models invokes heterogeneity of functional roles of individual components (*12*, *13*). It is reasonable to assume that the frequency of a component is mainly determined by the breadth of its functional role in the system. This explanation is especially applicable to biological and technological systems subject to natural and artificial selection respectively. Indeed, the frequency of genes whose "popularity" is not matched by their functional importance will be quickly corrected by the evolution. In agreement with this explanation, genes encoding certain core enzymes of central metabolism or ribosomal components are present in genomes of virtually all species (see Fig. 4 in Ref.(*14*)). On the other hand genes encoding peripheral enzymes tend to have much lower frequency of appearance in genomes (*14*). The same rule applies to multi-component software projects such as Linux where the most frequently installed components (e.g. "python", "gzip") are also among the most functionally important and reusable software libraries. Most other packages either directly or indirectly depend on these low-level components for their operation. As a result, these packages end up being installed on the vast majority (if not all) of individual Linux computers. In what follows we present empirical results supporting this second, functional explanation of the power law distribution of frequency of components in complex biological and technological systems.

## Results

### Empirical distribution of component frequencies

The eggNOG database (*15*) provided us with information about the presence or absence of genes from 45,000 orthologous gene families in genomes of more than 500 bacterial species. The Ubuntu popularity contest project quantified the frequencies of installation of about 200,000 Linux packages on more than 2,000,000 individual computers (*2*)(see Materials and Methods for details). We found the distributions of components' frequencies $f_i$ in both biological and technological systems to share multiple common features including a power-law scaling regime $P(f) \sim f^{-\gamma}$ with $\gamma \approx 1.5$ (Fig. 1B for genomes and Fig.1E for Linux) terminating with a peak at the maximal frequency $f \approx 1$ (Fig. 1A, D). This peak, formed by components present in the vast majority of systems, also manifests itself as a broad plateau at $f \approx 1$ in Zipf's rank-frequency plots (Fig. 1C,F). A broad distribution of gene frequencies has been previously reported in biological literature(*1, 16–19*). However, this study reports and explains its scaling exponent. U-shaped $P(f)$ distributions are sometimes plotted on semi-logarithmic scale(*1*) with piecewise linear fit used to define three types of components dubbed "core" ($f > 0.95$), "character" ($0.95 \geq f > 0.1$), and "accessory" ($f \leq 0.1$) genes(*16*). In Fig. 1A we validate these previous observations and demonstrate them for Linux systems (Fig. 1D). We also confirm the existence and explain the origins of a sharp crossover separating the core components with $f \approx 1$ from the rest of the distribution. We mathematically predict the number of core components to be around $\sqrt{N}$, where $N$ is the total number of components with non-zero frequencies which are functionally connected to the core. The empirical data are in approximate agreement with this prediction. The separation between character and accessory genes is less well defined. Indeed, when plotted in log-log coordinates the power law scaling observed for $f \ll 1$ directly crosses over into the core region at $f \approx 1$ without an obvious intermediate region corresponding to character genes. In what follows, we argue that the power law is expected on purely theoretical grounds. Thus "fractal organization of the gene Universe"(*20*) manifests itself both in the scale-free distribution of component frequencies (as reported in this study) as well as in qualitatively similar shapes of $P(f)$ at different evolutionary timescales (as demonstrated in Ref.(*1*)).

**Component's frequency is positively correlated with its dependency degree.**

It is reasonable to expect the frequency of a component (a gene or a software package) to be influenced by its importance or the breadth of its functional role in the system. For a given component, we quantify the latter by the number of other components whose operation critically depends on it either directly (referred to as the direct dependency degree $k_{dep}$) or directly+indirectly (referred to as the total dependency degree $K_{dep}$). The difference between $k_{dep}$ and $K_{dep}$ can be easily understood in the dependency network of Linux packages(*21*). Edges of this directed network connect a given package to packages it requests to install during its own installation process. Some of these packages have direct dependencies of their own. For example, Fig. S1 visualizes direct and indirect dependencies of the Firefox browser. This cascade of sequential installations continues until all downstream packages required for the operation of the chosen package are installed. So, while $k_{dep}(i)$ counts the packages that require installation of the package $i$ at the first step of this multi-step process, $K_{dep}(i)$ counts the packages that do so at any step.

While a similar interdependence of individual genes on each other certainly exists in biological systems it is more difficult to quantify. Using the algorithm described in Ref.(*13*) we calculated the dependency network for a subset of all gene families corresponding to metabolic enzymes (see(see Materials and Methods section for details).Briefly, our algorithm derives upstream-downstream relations of enzymes reflecting their relative positions in metabolic pathways. The functioning of an anabolic enzyme requires the presence of enzymes in the smallest pathway necessary to synthesize all of its substrates from the minimal set of core metabolites (see Materials and Methods section for our algorithm searching for such minimal pathway). The total dependency degree $K_{dep}(i)$ of the enzyme $i$ is given by the total number of enzymes in this minimal pathway located downstream from it for anabolic enzymes (or upstream from it for catabolic enzymes). On the other hand, the direct dependency degree, $k_{dep}(i)$ counts enzymes located one step below (or above) it in this hierarchy. The direct dependency degree of an enzyme is closely related to its degree in the adjacency matrix of the metabolic network previously studied in Refs. (*10*, *22*). Fig. S2 visualizes dependencies among enzymes in a particular metabolic pathway.

The scatter-plot of the frequency of a component versus its total (direct+indirect) dependency degree $K_{dep}$ clearly shows positive correlation between the two variables. The Spearman rank correlation 0.3(metabolic enzymes) and 0.47 (Linux packages) is highly statistically significant (p < $10^{-16}$). A somewhat weaker correlation for metabolic enzymes can be attributed to an important difference between dependency networks in biological and computer systems. The dependencies of software packages in Linux are explicitly specified by their designers and thus totally unambiguous. The biological systems are designed in a more robust fashion and allow some flexibility in dependencies among their components. For example, in metabolic networks there is often more than one enzyme synthesizing a metabolite used by another enzyme. This makes the definition of dependency degree of an enzyme more ambiguous and weakens its correlation with its frequency. To verify this hypothesis, we constructed dependency network of metabolites instead of metabolic enzymes and recomputed their usage frequencies in metabolic networks of different organisms (see Fig. S3). The correlation coefficient (0.45) was considerably better than for metabolic enzymes (0.3) and just slightly lower than that observed for Linux packages (0.47) (Fig 2).

**Dependency degrees follow power law distributions**

The distributions of direct ($k_{dep}$) and total ($K_{dep}$) dependency degrees for the metabolic enzymes as well as Linux packages are shown in Fig. 3. Both have a power-law scaling region with exponents around -2 ($k_{dep}$ shown in Fig. 3A) and -1.5 ($K_{dep}$ shown in Fig. 3B) correspondingly. In addition to the power law region Zipf's rank-degree plots of $K_{dep}$ (Fig. 4AB) but not of $k_{dep}$ have plateaus formed by the core components with the largest $K_{dep}$ (compare to frequency plateaus Fig. 1C,E). Direct dependency degrees in a variety of large software projects have been previously reported to have scale-free distribution with exponents around -2 (see Ref. (*23*) for Linux as well as in-degree exponents in table I of Ref. (*24*)).

## Discussion

One of the intriguing results presented above is a remarkable similarity of distributions of frequencies (Fig. 1) as well as topological properties of dependency networks (Fig 3) in biological (red circles) and technological (blue diamonds) systems. It is rather surprising to see near perfect overlap of distributions in these two systems of very different origins: one is optimized by Nature over billions of years of evolution, while the other is designed by a distributed population of human software engineers over the last several decades. In fact, below we argue that the functional form of $P(K_{dep})$ and $P(f)$ observed in this study is a universal property of any multi-component and multi-layered complex system. Such systems grow by gradually acquiring new components whose operation extends the functions performed by previously acquired components. Dependency networks connecting components to each other in such systems tend to be multi-layered as a direct consequence of the long history of growth and evolution (*25*). Metabolic and software dependency networks used in this study with 34 and >40 layers respectively are indeed multi-layered. A slightly different version of the universal metabolic network has been estimated (*25*) to have up to 60 layers of enzymes gradually acquired over billions of years of biological evolution (see Figs. 6,7 of Ref.(*25*) )

One mathematically tractable example of a multi-layered dependency network is provided by a critical random branching tree (*26*) viz. a tree with the branching ratio $b$ close to 1. Here the branching ratio $b \leq \langle k_{dep} \rangle$ counts nodes that directly depend on a given node and are located one layer above it. Indeed, in a branching tree with $b$ significantly larger or smaller than 1, either the number of layers is logarithmically small ($b \gg 1$) or the branches terminate prematurely ($b \ll 1$) rendering a multilayered network impossible.

For a critical branching tree one can show that $P(K_{dep}) \sim K_{dep}^{-\gamma}$ with $\gamma = 1.5$. Indeed, the part of the tree located upstream of a given node itself constitutes an instance of a critical branching process that is independent from the rest of the tree. Therefore, its size is distributed with the Galton-Watson exponent $\gamma = 1.5$ (see Ref. (*26*) for the mathematical derivation). Since no subtree can be larger than the parent tree, in a tree of size $N$ one expects to find $N \cdot P(K_{dep} \geq N) = N \cdot N^{1-\gamma} = \sqrt{N}$ subtrees with sizes about $N$. Therefore, about $\sqrt{N}$ nodes located at the lowest layers of the dependency network will have the largest possible dependency degree $K_{dep} \sim N$. (see Materials and Methods for more details). These nodes constitute the plateau in Zipf's plots (see Figs. 4A,B, 1C,E) and the large-x peak in the U-shaped distribution of dependency degrees or frequencies of system's components (see Fig. 1A,D).

While the distribution of dependency degrees in a critical branching tree is in excellent agreement with the empirically observed data, there is a conceptual difference between real-life dependency networks and trees. Indeed, in a tree each component directly depends on one and only one downstream component. On the other hand, in real-life networks this number, $D$, is certainly larger than one. It varies from component to component but on average tends to be around 2 for both metabolic networks and Linux packages. To describe real-life dependency networks with $D > 1$ we introduced and studied the following simple model. In our model dependency networks starts to grow from a few seed components. At each evolutionary time step one adds a new component depending on $D_i$ randomly selected existing components. For simplicity we assume $D_i$ to have a Poisson distribution with average $\langle D_i \rangle = D$. However, as shown in supplementary materials our results depend only on the average value of $D_i$ $D_i$. We mathematically derive (see supplementary materials for step-by-step calculations) that the total dependency degree $K_{dep}$ in a dependency network of size $N$ generated by this model has a power law tail $P(K_{dep}) \sim K_{dep}^{-(1+1/D)}$ as well as a plateau in the Zipf's plot composed of $N^{(D-1)/D}$ nearly universal components. Simulations of the model with $D = 2$ and $N = 1500$ (green line in Fig. 4A) provides a reasonable fit to the metabolic dependency degree distribution (blue diamonds in Fig. 4A), while $D = 2$ and $N = 10,000$ (green line in Fig. 4B) is an excellent fit to the Linux dependency degree distribution (red circles in Fig. 4B).

We see that the model with $D = 2$ provides a rather good fit to dependency networks in both biological and technological systems. Metabolic enzymes usually have two substrates and rarely one or three and more substrates. Hence in this case there is a good biophysical explanation for the observed value of $D_{met} = 1.7 \simeq 2$. The situation is more complicated for Linux dependency network where there are no geometrical limitations on the number of direct dependencies of a software package. This network is characterized by a large number of direct links between packages already indirectly connected on each other. Such shortcuts (known as feed-forward loops in the network jargon) do not change the overall (direct+indirect) network of package dependencies. Since our model does not contain feed-forward loops beyond those created by pure chance we pruned them from the Linux direct dependency network as well. After removing all direct links short-circuiting any chain of direct links in the Linux dependency network we were left with a direct dependency network with $D_{Linux} = 2.4 \simeq 2$ and the same set of direct+indirect package dependency links as the original network. Admittedly in the case of Linux packages we have no ready explanation for this particular value of $D$ beyond a vague notion that the easiest way to add package is to combine the outputs of two already existing ones.

The similarity between real-life dependency networks and those generated by our model with $D = 2$ extends beyond the shape of the total dependency degree distribution with the exponent $\gamma = 1 + 1/D = 1.5$ and $N_c = N^{(D-1)/D} = \sqrt{N}$ of the best connected components forming the plateau in Zipf's plots. Our model makes very specific predictions about how the dependency degree of a component depends on the time when it was first added to the dependency network. Unfortunately, obtaining system-wide information about these "creation" times is not easy for Linux and downright impossible for metabolic enzymes. As advocated in Ref.(*25*) the time of appearance of a metabolic enzyme in the metabolic pan-network can be estimated from its layer number obtained by the "scope expansion" algorithm. Using the layer number of a node in a real-life dependency network as a proxy of its acquisition/creation time we investigated its correlations with its total dependency degree. It stands to reason that "older" nodes located at bottom layers will tend to have systematically larger dependency degree in both model and real networks. This is indeed what was observed and shown Fig. 4C-E.

An important caveat in applying the $N_c = \sqrt{N}$ relationship is that $N$ counts only those components that are directly or indirectly connected to the core by the functional dependency network. For biological systems this allows to reconcile the apparent paradox. Indeed, the pan-genome of all bacterial species is believed to be open(*16*). That is to say, $N$ continues to increase without any hint at saturation as we sequence genomes of new bacterial species or even new strains of the same species (see e.g. Fig. 1 in Ref.(*27*)). At the same time the core bacterial genome remains relatively stable. Different methods result in somewhat different estimates of $N_c$ ranging from 250 in Ref. 16 to 400 in Refs.(*17–19*). To reconcile the apparent stability of $N_c$ with unlimited growth of $N$ one recalls that continuing expansion of $N$ is caused by either non-functional (prophages or transposable elements) or extremely niche-specific gene families. Both of them are likely to be disconnected from the core and hence will not contribute to growth of $N_c$. Assuming $N_c \leq 500$ one gets the upper bound on the number of gene families connected to the core at around 250,000.

The frequency of a given component is expected to be strongly correlated with its total dependency degree. Indeed, the system using any of $K_{dep}$ components located upstream of a given component is guaranteed to include this component itself. Hence, if every software package (metabolic enzyme) was equally likely to be initially selected (with probability $p_i = \frac{1}{N}$) for local installation on a computer (incorporation into a bacterial genome) one would have $f_i = \sum_{j=1}^{K_{dep}(i)} p_j = \sum_{j=1}^{K_{dep}(i)} \frac{1}{N} \sim K_{dep}(i)$. Deviations from this idealized linear relationship between $f_i$ and $K_{dep}(i)$ in real data reflect among other things a non-uniform frequency of initial selection or installation of upstream components. Indeed, idiosyncratic differences in popularity $p_i$ of higher-level components will be translated into differences in installation frequencies of lower-level components required for their operation. By adjusting the values of $p_i$ - the initial popularity of components - we were able to increase the correlation coefficient between $f_i$ and $\sum_{j=1}^{K_{dep}(i)} p_j \equiv \tilde{K}_{dep}(i)$ to 0.8 up from around 0.5.

Comparison between biological and technological networks has been previously performed in Ref.(*28*), and a number of similarities as well as significant differences was reported. However, the biological and technological systems studied by Yan et al. were rather different from the ones we used in this study. The focus of the analysis performed in Ref.(*28*) was on regulation and control represented by transcriptional regulatory network in *E. coli* and the call graph between subroutines within the Linux kernel. On the other hand, in this article we compare biological and technological systems with independently installable components represented by metabolic enzymes encoded in bacterial genomes and software packages installed on top of the Linux kernel. A more systematic analysis of similarities and differences between different versions biological and technological complex systems will have to await future studies.

## Materials and Methods

The methods are briefly summarized here while more detailed description is provided in SI Materials and Methods.

### Empirical data for frequencies of use of bacterial genes

The eggNOG database v3.0 (*15*) contains the mapping of orthologous gene families to 630 species with fully sequenced genomes. We included in our analysis 529 bacterial genomes and their gene families assigned based on the Clusters of Orthologous Genes (COGs) and universal Non-supervised Orthologous Groups (NOGs) which together cover 44283 prokaryotic orthologous gene families. The resulting table of presence or absence of individual gene families in genomes was then processed to obtain the gene frequency $f$ defined as the fraction of 529 genomes the family is represented by at least one gene.

### Empirical data for frequencies and mutual dependencies of Linux packages

The package dependency network of Linux distribution Ubuntu 11.04 Natty was obtained by first getting a complete list of packages from the web-page http://packages.ubuntu.com/, and then running the command *apt-rdepends* to find all the direct and indirect requirements for each package. The resulting network contains 33,473 packages, 157,667 direct, and 2,439,011 total (direct+indirect) dependency relations. The installation frequency data for 192,392 packages on 2,047,796 computers were downloaded from the package popularity contest project (http://popcon.ubuntu.com/by_inst)

### Construction of the dependency matrices for the metabolic network

We used the union of all reactions in the KEGG database(*29*)to construct upstream-downstream relations between enzymes using the following algorithm related to the "scope expansion" algorithm of Ref.(*25*). For every enzyme the minimal metabolic pathway connecting the product(s) of this enzyme to the set of five core metabolites was constructed as described in Ref. (*13*). The direct dependency links were then drawn between the selected enzyme and enzymes in the top layer of this pathway, while direct+indirect links connect it to all enzymes in the minimal pathway. The resulting dependency network contains 1832 reactions/enzymes connected to each other by 3118 direct and 49,168 direct+indirect dependencies.

### Power law fits to the data
Power law fits to distributions were performed using Matlab package plfit.m developed by Aaron Clauset and collaborators and downloaded from http://tuvalu.santafe.edu/~aaronc/powerlaws.

## Acknowledgements

We thank P. Dixit for useful discussions, critical reading and editing of the manuscript, K. Dill and J. Peterson for useful comments and suggestions. This work was supported by grants PM-031 from the Office of Biological Research of the U.S. Department of Energy.

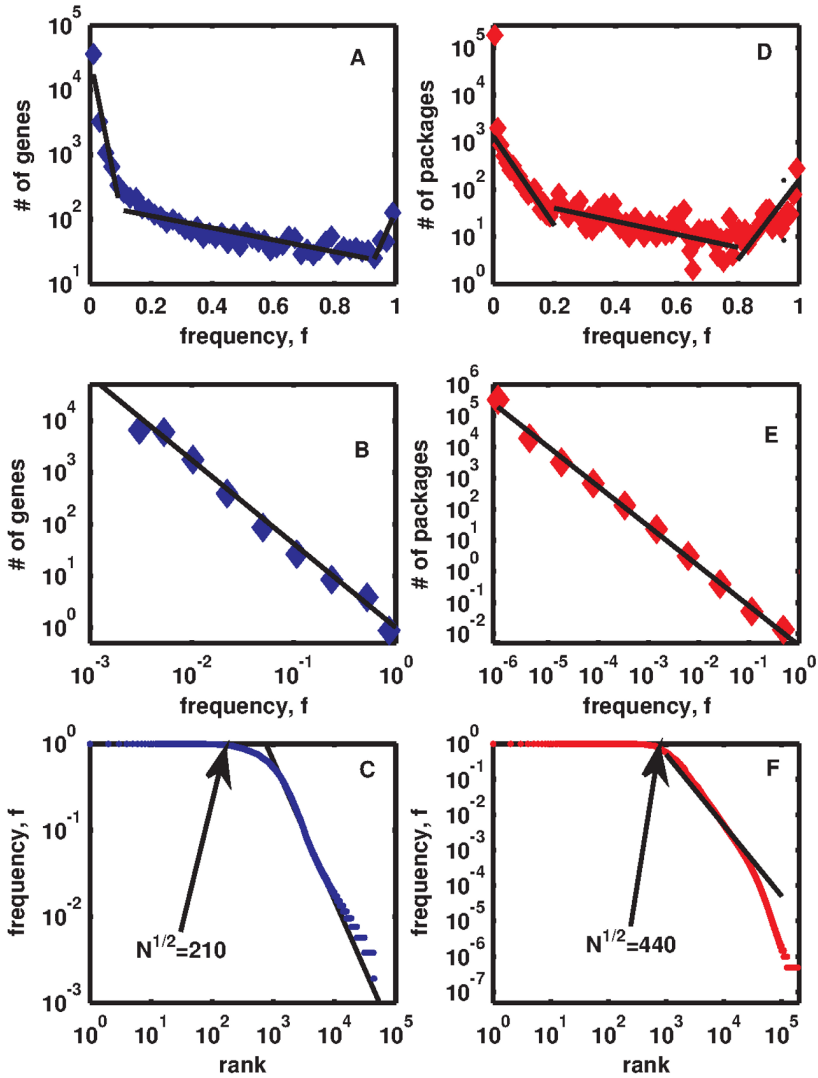

*Fig. 1. The histogram $P(f)$ of the frequency $f$ of bacterial genes present in genomes (panel A) or Linux software packages installed on computers (panel D) in semi-logarithmic coordinates. Dashed lines show a piecewise linear fit used to define "core" ($f > 0.95$), "character" ($0.95 \geq f > 0.1$), and "accessory" ($f \leq 0.1$) components(1, 16). When plotted in log-log coordinates (panel B for genes and E for Linux) the histogram is consistent with the power law $P(f) \sim f^{-\gamma}$ with the exponents $\gamma_{Genomes} = 1.62$, and $\gamma_{Linux} = 1.42$ (solid lines in panels B and E). In rank-frequency Zipf's plots (panel C for genes and panel F for Linux) core components manifest themselves as plateaus at $f \sim 1$. Straight lines in panels C and F are the best power law fits used to determine $\gamma_{Genomes}, \gamma_{Linux}$ and the arrows point to $\sqrt{N}$ - the mathematically predicted number of core components.*

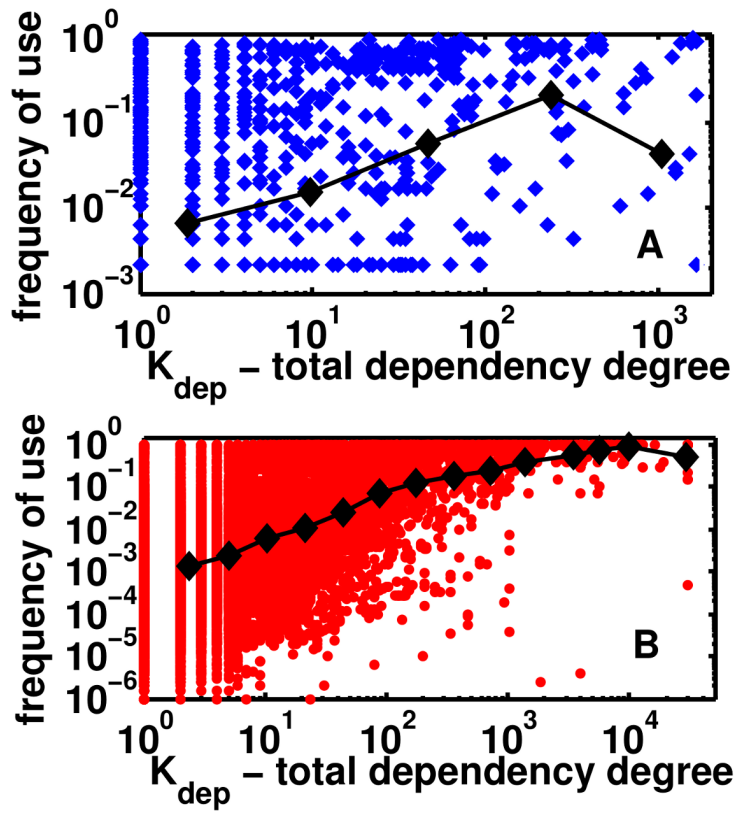

Fig. 2. Components' frequencies $f$ (y-axis) are positively correlated with their total (direct+indirect) dependency degrees $K_{dep}$ (x-axis) for both metabolic enzymes (panel A) (Spearman $r_s = 0.30$) and Linux packages (panel B) (Spearman $r_s = 0.47$). The black lines and symbols show the geometric averages of $f$ in each logarithmic bin of $K_{dep}$.

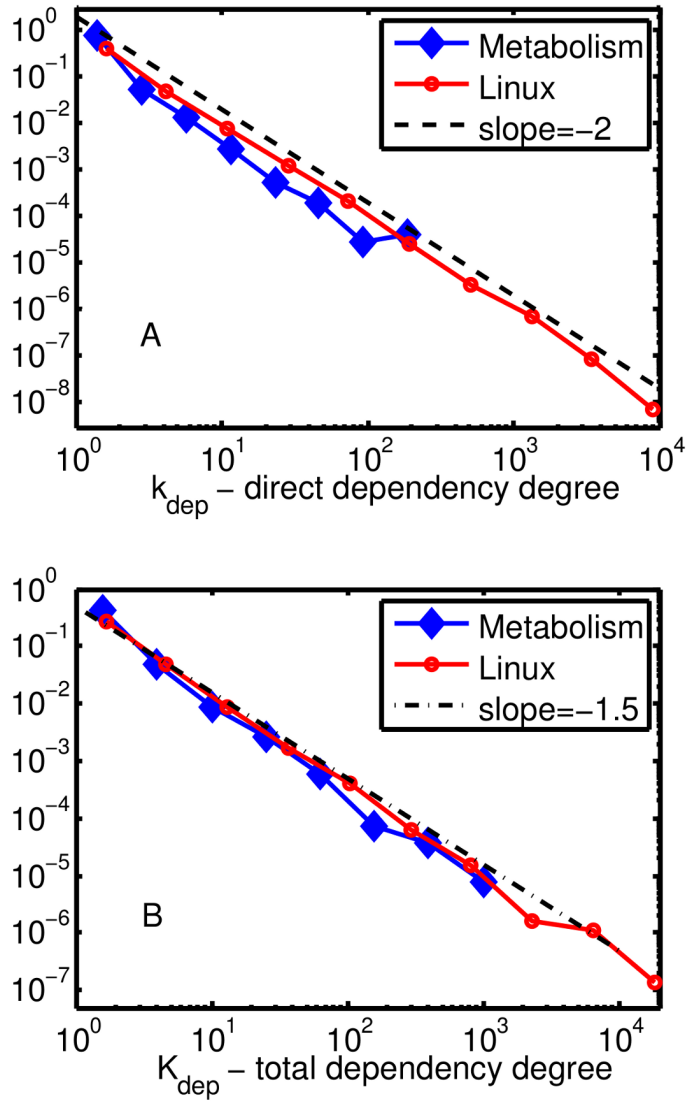

*Fig. 3. Probability distributions of direct ($k_{dep}$, panel A) and total ($K_{dep}$, panel B) dependency degrees for metabolic enzymes (blue diamonds) and Linux packages (red circles). Power law fits to direct degree cumulative distribution give -2.08 for metabolic enzymes and -1.91 for Linux packages, and are both consistent with the -2.0 scaling law (solid line in panel A). Power law fits to direct degree cumulative distribution give -1.5 for metabolic enzymes and -1.56 for Linux packages, consistent with the mathematically derived -1.5 scaling (solid line in panel B).*

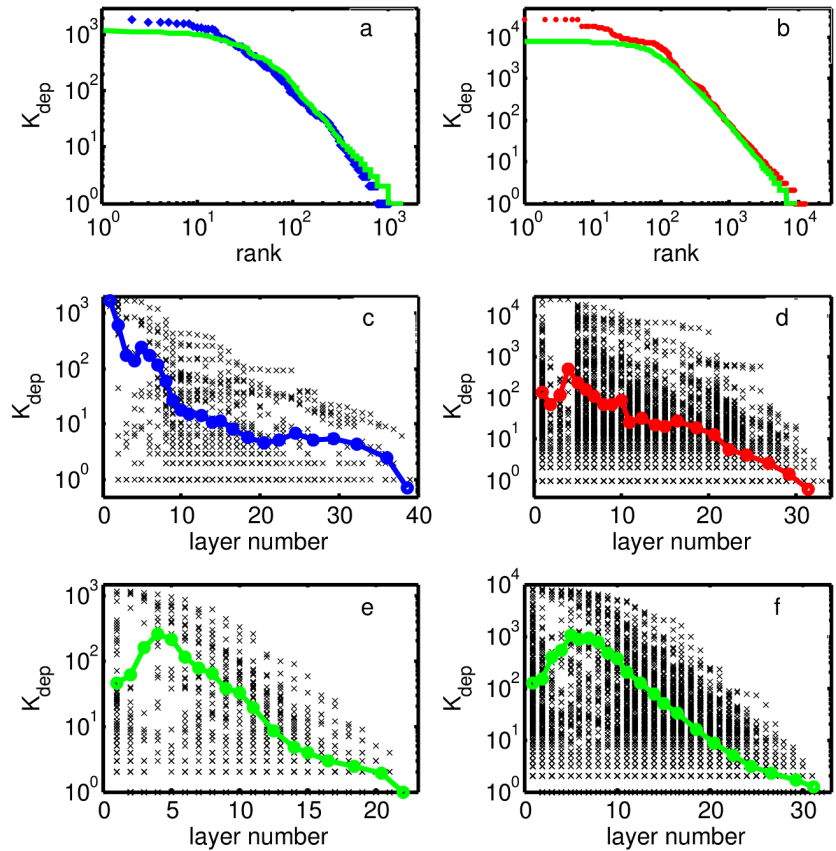

*Fig. 4. Zipf's plots of total dependency degree in real metabolic (blue symbols in Panel A) and Linux (red symbols in panel B) systems fitted with a random dependency model with D=2 and N=1500 (green symbols in panel A ) or N=10,000 (green symbols in panel B) respectively. Panels C and E show $K_{dep}$ vs. the layer number in the metabolic network and the best fitting random model respectively. Panels D and F do the same for Linux dependency network and its best approximation with random model. Black dots show scatter plots of individual nodes, while color lines are binned averages.*

# Supplementary Materials

## SI Materials and Methods

***Obtaining the dependency network and occurrence frequency data for Linux packages.***

The package dependency network of Linux distribution Ubuntu 11.04 Natty was obtained by first getting a complete list of packages from the web-page http://packages.ubuntu.com/natty/, and then running the command apt-rdepends to find all the direct and indirect requirements for each package. The resulting network contains 33,473 packages, 157,667 direct, and 2,439,011 total (direct+indirect) dependency relations.

The occurrence frequency data for 192,392 packages on 2,047,796 computers were downloaded from the package popularity contest (popcon) project (http://popcon.ubuntu.com) in our analysis we used the file listed under "statistics for the whole archive sorted by the field" → Inst (Institutions) (1). Participants of this project installed on their Linux computers tracking software that automatically reports the installation and subsequent usage of different packages to the popcon server. We used the first column reporting the number of computers where this package was installed. 189,711 packages were installed on at least one computer. Other columns not used in this study report the number of computers where this package was or was not used in the past month and the number of computers where it was recently updated.

The popcon project obtained the package data from Ubuntu Linux of a wide range of versions and CPU architectures, whose package repertoire and dependencies are a little bit different from each other. In the analysis we assumed that all participants are using Ubuntu 11.04 with x86 architecture, and based on this version of the Ubuntu Linux we calculated the direct and total dependency degree ($k_{dep}$ and $K_{dep}$) of every package, and plotted the $f$ vs. $K_{dep}$ in Fig. 2. The packages not included in the official repositories of Ubuntu 11.04 or having zero installation frequency were ignored. Packages that are not required by any other packages (i.e. those with $K_{dep} = 0$) were also ignored.

***Construction of the dependency matrices for the metabolic network***

The KEGG database (2) contains the data of metabolic reactions present in different organisms, and the universal metabolic network used in this study is the union of all the reactions in KEGG consisting of 5759 reactions and 4785 metabolites. The group of 5 common metabolites present in the majority of organisms was selected as the core: $H_2O$, ATP, NAD+, oxygen and Coenzyme A. The final version of the dependency network used in our study contains 1832 reactions (or associated enzymes) connected to each other by 3118 direct and 49,168 direct+indirect dependencies.

The goal of the metabolic network is to either convert nutrients taken up from the environment into core metabolites (catabolism), or to convert core metabolites into the constituents of the biomass and other essential ingredients (anabolism). The direction of the dependency network connecting metabolic reactions would be opposite in these two cases. For simplicity we will concentrate on the case of anabolic pathways below. For catabolic pathways we simply inverted the direction of reactions and then applied the procedure used for anabolic pathways.

In order to determine the set of other enzymes an enzyme $i$ in an anabolic pathway depends on for its operation we performed the following computational analysis. We selected all metabolic substrates of the enzyme $i$ one-by-one and for each of them we constructed the minimal pathway necessary to synthesize this metabolite from our pre-determined set of forty core metabolites. The union of all enzymes in these pathways constructed for each of the substrates of the enzyme $i$ is a good approximation to the minimal set of enzymes necessary to enable the reaction catalyzed by the enzyme $i$. As such we can plausibly assume that the enzymes in this union form the total downstream dependency set for the enzyme $i$.

Furthermore, by analogy to software dependency networks, the direct dependency neighbors of the enzyme $i$ are made by the set of enzymes added at the last layer of our breadth-first search algorithm. Based on this definition the direct dependency degree of an anabolic enzyme is closely related to the number of metabolic reactions using at least one of its products which is one of the standard topological definitions of degree in metabolic networks (3). Therefore, the power law distribution with the exponent -2 we measured for direct dependency degrees is closely related to previously reported scale-free topology on metabolic networks (3).

The rules by which this minimal pathway was constructed were previously described in Ref. (4). For the sake of completeness we included them in the supplementary text below.

By repeating the above procedure for all anabolic (catabolic) enzymes located downstream (upstream) from our core metabolites and thus reachable from the core by the scope expansion algorithm (5) we constructed our best approximation to the total dependency network of metabolic enzymes in the KEGG database.

*Rules of addition of anabolic pathways in dependency network calculation.*

1. At the beginning of the simulation, the model organism starts with a "seed" metabolic network consisting of 5 metabolites including $H_2O$, ATP, NAD+, oxygen, and CoA. It is assumed that our organism is able to generate all of these metabolites by some unspecified catabolic pathways.

2. At each step a new metabolite that cannot yet be synthesized by the organism is randomly selected from the "scope" (5) of our seed metabolites. This scope consists of all metabolites that in principle could be synthesized from the seed metabolites using all reactions listed in the KEGG database (see Ref. (5)).

3. To search for the minimal pathway that converts core metabolites to this target we first perform the "scope expansion" (5) of the core until it first reaches the target. In the course of this expansion reactions and metabolites are added step by step (or layer by layer). Each layer consists of all KEGG reactions that have all their substrates among the metabolites in the current metabolic core of the organism (light blue area in Figure 4 of Ref. (5)) and those generated by reactions in all the previous layers. (See Figure 4 of Ref. (5) for an illustration).

*Mathematical derivation of the total dependency degree distribution in the random model with $D=2$.*

To mathematically derive the distribution of dependency degree $K_{dep}$ in the simple model proposed in this study we study its dependence on the time $t$ a package was added to the growing dependency network. Here time $t$ is defined as the size of the network when a package was added and may have a non-linear but monotonic relation to the actual time of addition (e.g. in exponentially expanding systems). $K_{dep}(t)$ can be calculated self-consistently from the following equation:

$$K_{dep}(t) = 1 + \int_{t+1}^{N} K_{dep}(t')D/t' \tag{1}$$

Indeed, the total dependency degree of a package added at time $t$ is given by the sum of total dependency degrees of packages added at later times $t'$ that directly depend on it Indeed $K_{dep}$ counts both direct and indirect dependencies and thus indirect dependencies of upstream packages are transferred to their downstream neighbors. In a random model the likelihood of a package added at time $t'$ to send a direct dependency link to a package added at time $t$ is simply $D/t'$. It is easy to check that

$$K_{dep}(t) = (t/N)^{-D} \tag{2}$$

is a solution of this equation. Indeed,

$$1 + \int_{t+1}^{N} D\left(\frac{t'}{N}\right)^{-D} dt'/t' \approx 1 + (\frac{t}{N})^{-D} - (\frac{N}{N})^{-D} = (\frac{t}{N})^{-D} = K_{dep}(t)$$

. The equation (1) simply adds up the dependency degrees of multiple upstream neighbors of a node and thus ignores the inevitable overlap between these sets of nodes. This is a good approximation as long as the resulting $K_{dep}(t) \ll N$ and thus the overlap is small. It is clear however that if $D > 1$ the equation (2) cannot hold forever since it predicts $K_{dep}(1) = (1/N)^{-D} = N^D \gg N$. The total dependency degree cannot be larger than $N$ and this value is approximately reached at $t = N_c$ determined by $(N_c/N)^{-D} = N$ or

$$N_c = N^{(D-1)/D} \tag{3}$$

$N_c$ is the number of nearly universal "core" components in the system with total dependency degree $K_{dep} \approx N$.

Eq. (2) fully determines the power law tail of the distribution of dependency degrees. Indeed, $P(K_{dep} \geq K) = P((t/N)^{-D} \geq K) = P((t \leq NK^{-1/D}) = NK^{-1/D}/N = K^{-1/D}$. Hence, $P(K_{dep} = K) = -dP(K_{dep} \geq K)/dK$ is given by

$$P(K_{dep}) \sim K_{dep}^{-(1+1/D)} \tag{4}$$

For $D = 2$ which is close to its empirical value in real-life biological and technological systems used in this study one recovers familiar scaling laws:

$$P(K_{dep}) \sim K_{dep}^{-1.5} \tag{5}$$

and

$$N_c = \sqrt{N} \qquad (6)$$

***Mathematical derivation of the total dependency degree distribution in a tree generated by a Galton-Watson branching process.***

A Galton-Watson branching process is a Markov process in which every node in generation $l$ produces some random number of "child nodes" in generation $l+1$, according to a fixed probability distribution that does not vary from node to node. We denote as $p_0$ the probability for the process to terminate at each node, while $p_d$ is the probability for a node to have a branch with $d$ child nodes. The first node of the tree generated by a Galton-Watson branching process is denoted as the root. In biological and technological systems considered in this study the root node represents the set of core metabolites, or the basic Linux packages that serve many high-level user applications. The scaling properties of the Galton-Watson process are fully determined by a single parameter $\overline{d} = \sum_{d=0}^{d_{max}} d \times p_d$, which is the average number of child nodes of any given node has. For $\overline{d} < 1$, referred to as under critical branching process, the cascade will terminate very quickly and is irrelevant to this study. Conversely for $\overline{d} > 1$, referred to as supercritical branching process, the cascades will likely never terminate. Moreover, for a given number of nodes $N$ in a tree generated by an overcritical branching process, the total number of layers $L$ is logarithmically small: $L \sim \log N / \log \overline{d}$. Real-life complex multi-component systems such as e.g. metabolic networks and large software projects are characterized by a large number of hierarchical levels (4) incompatible by that in an overcritical branching process. Thus overcritical branching processes will be also ignored in this study. In what follows we will limit our calculations to the third case in which $\overline{d} = 1$ denoted as critical branching process. This was previously demonstrated to be a good approximation to universal metabolic network (4) and the present study presents convincing evidence that it describes large software projects as well.

The direct dependency degree $k_{dep}$ of a node in the Galton-Watson process is given by its number of child nodes plus one (to account for the dependency of a node on itself). The distribution of $k_{dep}$ is then determined by $p_d$ as $P(k_{dep} = 1) = p_0$, $P(k_{dep} = 2) = p_1$, ... $P(k_{dep} = d) = p_{d-1}$. It can have any functional form as long as its average is equal to $1 + \overline{d}$, that for a critical branching process is equal to 2. It's important to emphasize that Galton-Watson branching process does not provide an explanation for the power law form of the distribution of direct dependency degrees (see Fig. 3). On the other hand, the total dependency degree $K_{dep}(i)$ corresponds to the size of the entire sub-tree initiated at the node $i$. The Galton-Watson branching process is a Markov process and thus each node can be thought as starting its own instance of a branching process that is independent of branching ratios of its predecessors. Hence, one would naively expect that the total dependency degree of $N$ nodes in a tree generated by the critical branching process will have the same power law distribution $P(K_{dep}) \sim K_{dep}^{-1.5}$ as $N$ independently started branching processes. However, a quick calculation convinces one otherwise. Indeed, the largest dependency degree $D_{max}$ in this case will be determined by the equation $1/N = P(K_{dep} > D_{max}) = \sum_{D_{max}} K_{dep}^{-1.5} \sim D_{max}^{-0.5}$, or $D_{max} = N^2$. Thus the dependency degree cannot be larger than $N$ - the total number of nodes in the tree. The size of the universal network with $N$ nodes imposes a strict cutoff of $N$ on sizes of its subtrees. Thus the following process reproduces the distribution of sizes of subtrees of a critical branching tree with $N$ nodes. In this process one simulates the critical branching process $N$ times and stops it when and if its size $s$ reaches $N$ nodes if it does not terminate on its own before that. Therefore, among $N$ nodes of the critical branching tree one expects to find $N \times P(s \geq N) = N \times N^{-0.5} = \sqrt{N}$ nodes with the largest total dependency degree $K_{dep} = N$. The rest of the nodes follow the power law distribution $P(K_{dep}) \sim K_{dep}^{-1.5}$. This is indeed what we see in our numerical simulations on the universal network of 5000 nodes generated by the critical branching process with $p_0 = p_2 = 1/2$. The Zipf's plot of total dependency degrees in this system is shown as green symbols in Figure 3D. The predicted $\sqrt{N} = 71$ for the number of (nearly) universal core components is in good agreement with the crossover between the plateau and power law regimes of Zipf's plot.

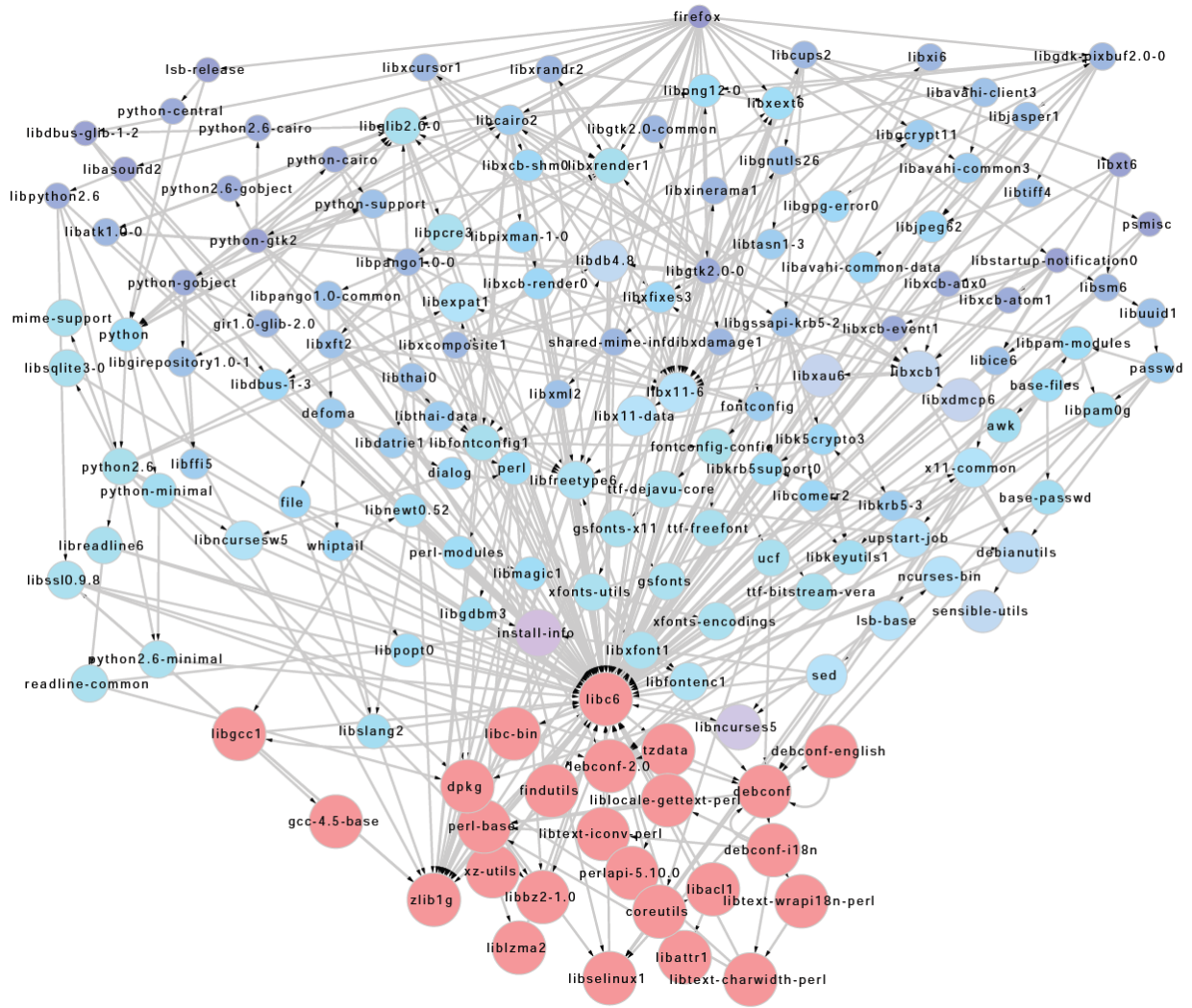

Fig. S1. Pathway diagram created with Cytoscape (6) showing 153 Linux packages directly or indirectly required for installation of the package firefox for the Firefox web browser (the top node). Size and color of the nodes correspond to their total dependency degree $K_{dep}$ (red-large and blue-small)

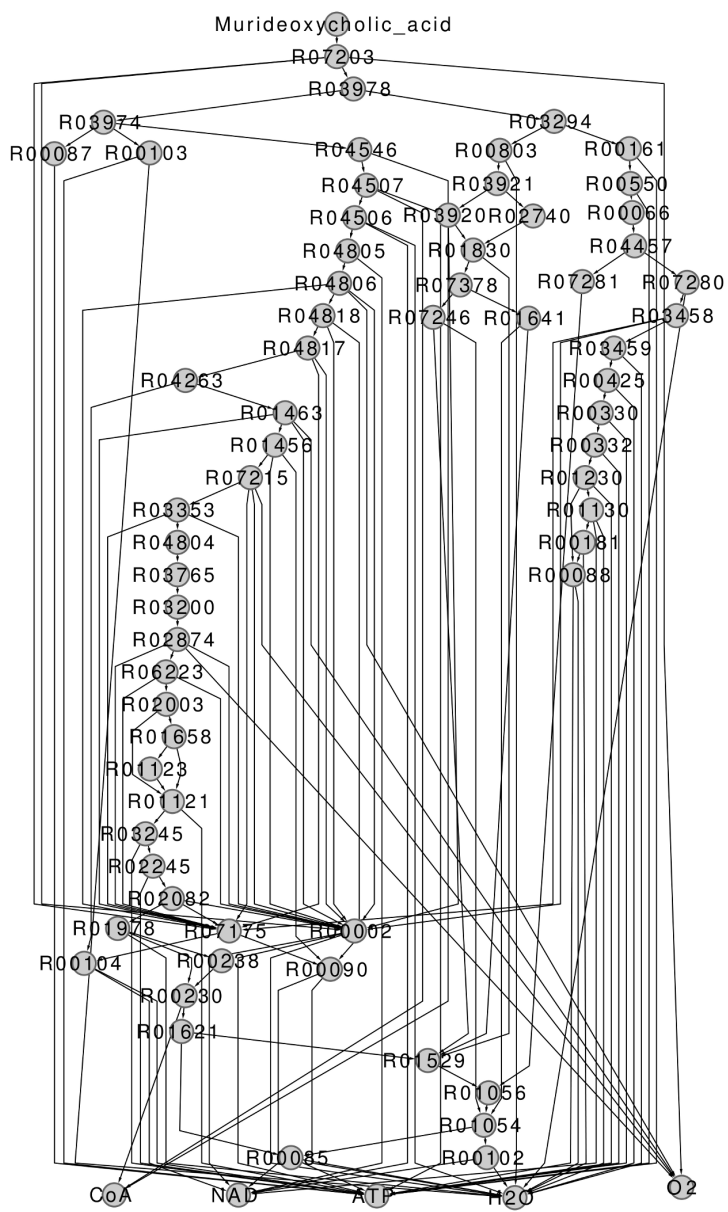

Fig. S2. Pathway diagram created with Cytoscape (6) showing 65 reactions (or equivalently enzymes catalyzing these reactions) that the production of the metabolite Murideoxycholic acid (C15515) in the KEGG database (the top node) directly or indirectly depends on. Reactions numbers are given in KEGG notation.

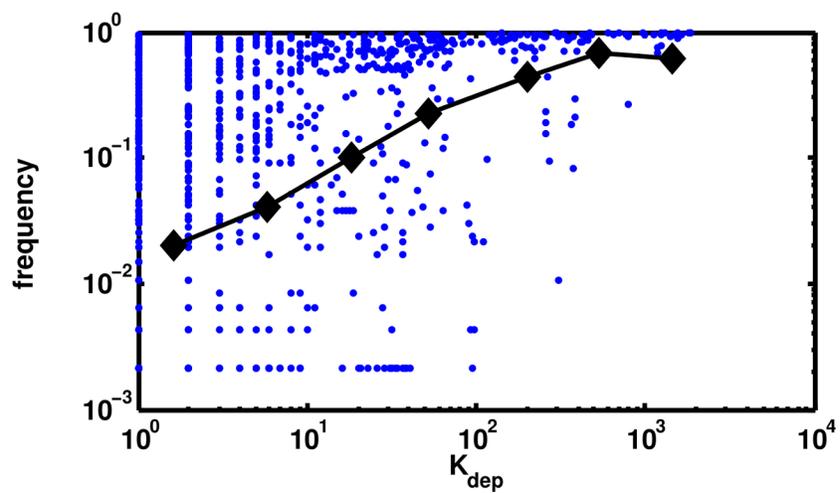

Fig. S3. The frequency of occurrence $f$ (y-axis) vs. the total (direct+indirect) dependency degree of metabolites $K_{dep}$. The two quantities are positively correlated (*Spearman* $r_s = 0.45$). The black curve and symbols shows the average $f$ in the logarithmic bins of $K_{dep}$.